\documentclass[conference]{IEEEtran}
\AtBeginDocument{%
  \providecommand\BibTeX{{%
    \normalfont B\kern-0.5em{\scshape i\kern-0.25em b}\kern-0.8em\TeX}}}

\usepackage{ifthen, xspace, float}
\usepackage{caption}
\usepackage{subcaption}
\usepackage{soul}
\usepackage{amsmath,amssymb}
\usepackage{enumitem}
\usepackage{graphicx}
\usepackage{url}
\usepackage{etoolbox, xstring}
\usepackage{booktabs}
\usepackage{array}
\usepackage{pifont}
\usepackage{siunitx}
\usepackage{svg}
\usepackage{multirow}
\usepackage{tablefootnote}
\DeclareListParser{\doslashlist}{/}
\newcounter{ndnNameComponentCounter}%
\newcommand{\name}[1]{{%
		\setcounter{ndnNameComponentCounter}{0}%
		\renewcommand{\do}[1]{{%
				\ifnumgreater{\value{ndnNameComponentCounter}}{0}{\allowbreak/}{}%
				\ifnumodd{\value{ndnNameComponentCounter}}{}{}%
				\detokenize{##1}}%
			\stepcounter{ndnNameComponentCounter}}%
		``{\fontfamily{cmtt}\small\selectfont\IfBeginWith{#1}{/}{/}{}\doslashlist{#1}}''%
}}

\newboolean{showcomments}
\setboolean{showcomments}{true}
\ifthenelse{\boolean{showcomments}}
{ \newcommand{\mynote}[3]{
    \protect\fbox{\bfseries\sffamily\scriptsize#1}
    {\small$\blacktriangleright$\textsf{\emph{\color{#3}{#2}}}$\blacktriangleleft$}}}
{ \newcommand{\mynote}[3]{}}

\usepackage{xcolor}

\definecolor{uclablue}{rgb}{0.33, 0.41, 0.58}

\newcommand{\etc}{etc.\@\xspace}

\newcommand{\eg}{\textit{e.g.,}\@\xspace}
\newcommand{\ie}{\textit{i.e.,}\@\xspace}

\def\second{({ii})\xspace}

\definecolor{verylightgray}{gray}{0.8}

\begin{document}
\title{Exploring the Design of Collaborative Applications via the Lens of NDN Workspace}

\author{\IEEEauthorblockN{Tianyuan Yu}
 \IEEEauthorblockA{\textit{UCLA} \\
USA \\
tianyuan@cs.ucla.edu}
\and
\IEEEauthorblockN{Xinyu Ma}
\IEEEauthorblockA{\textit{UCLA} \\
USA \\
xinyu.ma@cs.ucla.edu}
\and
\IEEEauthorblockN{Varun Patil}
\IEEEauthorblockA{\textit{UCLA} \\
USA \\
varunpatil@cs.ucla.edu}
\and
\IEEEauthorblockN{Yekta Kocaogullar}
\IEEEauthorblockA{\textit{UCLA} \\
USA \\
ykocaogullar@g.ucla.edu}
\and
\IEEEauthorblockN{ Lixia Zhang}
\IEEEauthorblockA{\textit{UCLA} \\
USA \\
lixia@cs.ucla.edu}
}

\maketitle
\begin{abstract}
Metaverse applications desire to communicate with semantically identified objects among a diverse set of \emph{cyberspace entities}, such as cameras for collecting images from, sensors for sensing environment, and users collaborating with each other, all could be nearby or far away, in a timely and secure way.
However, supporting the above function faces networking challenges.
Today's metaverse implementations are, by and large, use secure transport connections to communicate with cloud servers instead of letting participating entities communicate directly.
In this paper, we use the design and implementation of \emph{NDN Workspace}, a web-based, multi-user collaborative app to showcase a new way to networking that supports many-to-many secure data exchanges among communicating entities \emph{directly}.
NDN Workspace users establish trust relations among each other, 
exchange URI-identified objects directly, and can collaborate through intermittent connectivity, all in the absence of cloud servers. 
Its data-centric design offers an exciting new approach to metaverse app development. 
\end{abstract}




\section{Introduction}
In his book ``The Metaverse: And How It Will Revolutionize Everything'', the author and investor Matthew Ball envisioned the metaverse as --
\begin{quote}
\textit{A massively scaled and interoperable network of real-time rendered 3D virtual worlds that can be experienced synchronously and persistently by an effectively unlimited number of users with an individual sense of presence, and with continuity of data, such as identity, history, entitlements, objects, communications, and payments.}
\end{quote}
To create a shared, persistent virtual world, metaverse applications need to gather data from multiple sources (sensors, actuators, users, computing devices etc.) \emph{timely} and \emph{securely}.  Ideally, all communications should happen directly between users and/or devices to minimize latency, and with fine granularity access control at the level of individual media objects to allow ``owners'' to retain full control over their virtual objects.

However, the realization of this grand vision faces networking challenges today. 
The existing metaverse implementations largely rely on cloud servers -- all entities, both users and devices, talk to each other by setting up secure connections to cloud servers to exchange data.  
This model of communication has at least two shortcomings.
First, the long communication paths (from entity $A$ to the cloud server to entity $B$) introduce unnecessary delays.
Second, cloud servers are owned and controlled by their respective service providers, who also possess and control users' data and their access, resulting in centralization of the control power over user data.

To understand why today's state of affairs in networking is in such an undesirable situation, we look back the original Internet design: the goal was to support \emph{direct} communication from a host to any other host on the Internet, and the Internet indeed operated in this model during its early years.
Over time, however, direct end host communications have been replaced by today’s practice of communicating through the cloud. 
We attribute this shift to two closely related causes: the shortage of IP addresses which resulted in the loss of direct reachability between end hosts, and the necessity of \emph{secure} communications that TCP/IP cannot offer by itself.

In this paper, we use the development of a specific application, \emph{NDN Workspace}, as a case study to explore a data-centric approach.
\emph{NDN Workspace} utilizes semantically meaningful names for both users and data \footnote{Names that are human-recognizable carry semantic meaning, as they allow the identification of the named entity. In this paper, we refer to these human-recognizable names as semantically meaningful names, or semantic identifiers.}, and uses named, secured, and immutable data as the basic building block. 
This data-centric design enables direct, secure, and timely communication among multiple entities through any available connectivity, as well as asynchronous communication via in-network storage.

This paper makes two contributions.
First, NDN  Workspace showcases a new direction to networking, which facilitates the realization of the metaverse vision.
Second, the supporting libraries we have developed in realizing NDN Workspace can be directly used to facilitate new metaverse application developments.

The rest of the paper is organized as follows. 
Section \ref{overview} offers a brief overview of NDN Workspace, followed by a detailed description of its design in \ref{building}.
Section \ref{imple} describes the NDN Workspace implementation, followed by a quick evaluation in Section \ref{evaluation}. 
Section \ref{discuss} discusses several issues in building metaverse applications over Named Data Networking (NDN)~\cite{ndn14}. Section \ref{conclude} concludes the paper.

\section{NDN Workspace Overview}
\label{overview}
In this section, we first give an overview of the NDN Workspace application, then identify the necessary components to satisfy its design requirements, and provide a brief introduction to the key enabler --- Named Data Networking~\cite{ndn14}.

NDN Workspace is a collaborative application which is built into browsers to ease end user deployment.
A group of users who want to jointly develop \textit{a set of documents} form a \emph{workspace instance}, in which they jointly edit shared documents through direct user-to-user data exchanges
In this paper, we use ``NDN Workspace'' to mean the application we developed, and ``workspace'' for each specific running instance.
\begin{figure}[ht]
    \centering
    \includegraphics[width=0.45\textwidth]{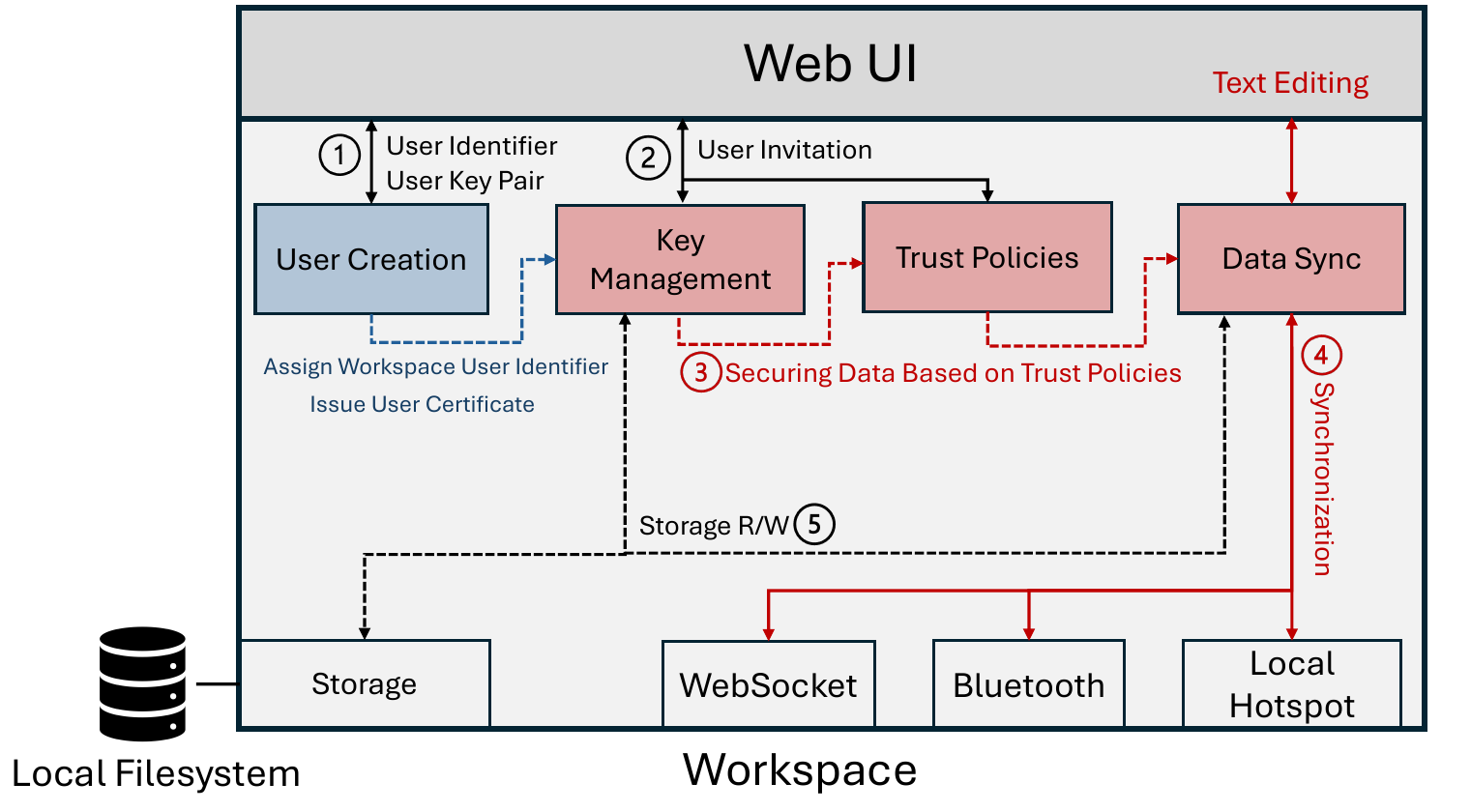}
    \caption{This figure shows the NDN Workspace application architecture. Users access the application through a browser. The browser takes a user's identifier and goes through security process to obtain the user's certificate and trust policies. 
After the bootstrapping step, the browser can name and secure all text edits the user makes, disseminates the data to other participants, as well as verify all the received data produced by others, and stores data into local disk.}
    \label{fig:app-structure}
\end{figure}

As Fig.~\ref{fig:app-structure} shows, each user obtains a \emph{semantically meaningful} identifier within a workspace instance.
Semantically identified users can bring their out-of-band trust relations into the application and endorse each other by issuing certificates to each other.
NDN Workspace names data instead of data containers, thus it can utilize any and all available connectivity (\eg WiFi, Bluetooth, Cellular, \etc) to communicate.
Following a data-centric approach, Workspace names data by URI-like identifiers, which are independent from nodes (data containers) or communication channels.
Every piece of data is encrypted and signed by its producer's key, which enables communications among users directly.
Receivers can validate all incoming data based on application specific security policies defined by the users, so that each workspace only consumes data that its users' trust policy allows. 
Moreover, users can continue to work on shared files even when they are offline, i.e. without Internet connectivity.
All the local changes a user $U$ makes to the shared files will be synchronized when $U$ gets connected with other users sharing the same workspace.

\vspace{-2pt}
\subsection{Semantic Identifier Is The Key}
NDN Workspace secures data directly using user-defined security policies.
Thus we need a systematic approach to specify the policies about \textit{who} can produce data under \textit{which} data namespace, and \textit{who} can consume data under \textit{which} namespace.
This requires assigning structured and semantically meaningful identifiers to users and data, so that schematized security policies can be defined on semantic, hierarchical namespaces. 
For example, a simple security control policy may state that only the user with identifier \name{alice@example.com} can produce data under the identifier prefix \name{yourworkspaces.app/MeetRoom/alice@example.com}.
Semantically identified and secured (further discussed in \S\ref{ndn}) data can be exchanged over any physical connectivity, and saved at any place with storage space; consumers can verify all received data independent from data containers or transmission channels.

\subsection{Background: Named Data Networking (NDN)}
\label{ndn}
Our data-centric approach to the NDN Workspace design follows the design direction of NDN~\cite{ndn14}. 
Today's TCP/IP Internet model views a network as made of interconnected nodes which are identified by IP addresses; communication security is an afterthought that is patched onto each end-to-end connection, e.g. running TLS on top of TCP. 
NDN, on the other hand, views a network as made of \emph{semantically named} entities with various \emph{trust relations} among each other~\cite{2021plug-play}; these named entities can be users, devices, services/app instances, or anything that produces and/or consumes named data.
NDN can directly use application layer data identifiers in network communications, without having to mapping application layer identifiers (e.g. DNS names) to IP addresses for packet delivery as the Internet does today.

In an NDN connected world, data consumers request data by sending \emph{Interest} packets with semantic names, and in response, the network returns the requested \emph{Data} packets.
A Data packet carries the matching semantic name, content and a cryptographic signature by its producer, which consumers use to authenticate the received packet.
\begin{figure}[h]
    \centering
    \includegraphics[width=0.48\textwidth]{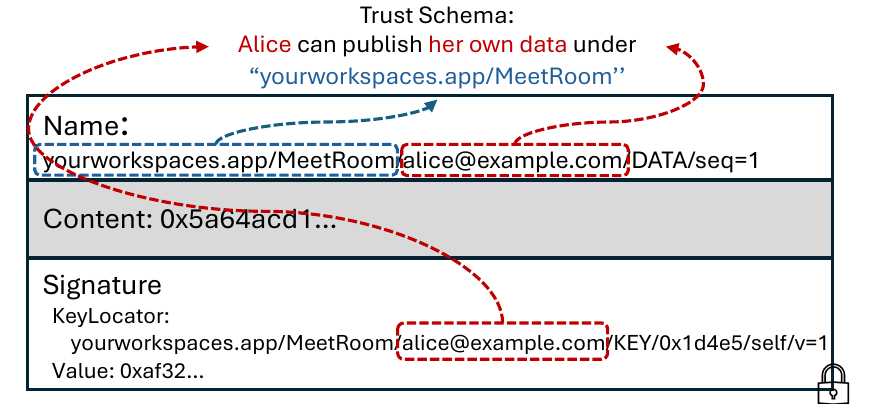}
    \caption{Alice's Data packet includes its packet name, content, and signature. The signature field includes a key locator, which refers another Data packet from Alice that contains her key.} 
    \label{fig:data-packet-example}
\end{figure}
To secure data piece directly, each NDN entity needs to go through a bootstrapping~\cite{yu2021enabling} process first to obtain its name, certificate, trust anchor and security policies. This step is remotely analogous to bootstrapping a host $H$ to the Internet, allowing $H$ to obtain its IP address and a few other necessary parameters needed to send and receive IP packets. 
The security policies are defined by applications and expressed as a set of schematized trust rules, called \emph{trust schema}~\cite{yu2015schematizing} and written in domain-specific languages~\cite{nichols2021trustschemas, yu2023trustschema}.
They define which Data producer's key, which is identified by its semantic name, is allowed to sign the Data packets of given names.
Fig.~\ref{fig:data-packet-example} depicts an example Data packet Alice produced under the namespace \name{yourworkspaces.app/MeetRoom}.
Data consumer can define trust policies to only accept \name{MeetRoom} Data signed by the same producer referred in the Data name.

Data confidentiality can be supported by content encryption. 
NDN has developed a specific solution, Named-based Access Control (NAC)~\cite{2018nac} scheme. In addition to encrypting data, a unique feature of NAC is \emph{automated} encryption and decryption key management which follow the data access policies defined in application-specific trust schemas.
NDN treats \emph{all} cyberspace objects, including app contents of all types (binary, image, video, etc.), cryptographic keys, and security policies, as semantically named and secure data. Therefore, crypto keys and trust schemas can all be fetched by their names in the same way as any other types of data.

Finally, multiple producers can contribute data to the same application continuously. 
NDN enables consumers to learn the names of the latest data production by its transport service, Sync (short of synchronization), whose job is to synchronize the names of all the shared data among all participants in the same application group~\cite{syncsok}. Whenever a producer in a ``Sync group'' produces a piece of new data, Sync notifies all other participants about the data name; individual entities can then decide whether and when to fetch the desired data. 
An offline entity can catch up when it reconnects and learns about the names of new data it missed.

\section{Developing NDN Workspace}
\label{building}
In this section, we show how NDN Workspace is realized by using semantically named and secured data as its basic building blocks.
We first explain how one can initiate an NDN workspace instance (\S\ref{workspace-identifier}), and how NDN Workspace assigns semantic identifiers to users (\S\ref{user-identifier}), followed by data namespace design (\S\ref{data-identifier}) and NDN Workspace membership management (\S\ref{authorization}).
We then explain the security workflow for data production and consumption (\S\ref{prod-cons}), and how NDN Workspace synchronizes secured, immutable data publications to achieve consistent document views among all users (\S\ref{sync}).
Finally, we talk about how NDN Workspace supports asynchronous collaboration utilizing in-network storage (\S\ref{async}).

\subsection{Semantic Workspace Identifiers}
\label{workspace-identifier}
A user can create a workspace as its initiator by first requesting a name for it; one can get a DNS name from a local DNS domain operator (\eg company.com, or cs.univ.edu) or directly from a DNS registrar (\eg GoDaddy).
Given DNS names are unique, this provides each workspace instance a unique semantic identifier on the Internet.
For example, Bob may own a DNS name \name{yourworkspaces.app} and obtain a certificate for \name{yourworkspaces.app}.
Under that name, Bob can initiate a new workspace instance with that a workspace named \name{yourworkspaces.app/MeetRoom} (we use \name{MeetRoom} for short in the rest of the paper)
, and starts the Workspace application.
Workspace creates a key pair for \name{MeetRoom} and signs it with the domain-certifying key (\eg X.509 certificate or DNSSEC key) of \name{yourworkspaces.app}.

Initiating his workspace, Bob is able to get a user identifier in \name{MeetRoom} (\S\ref{user-identifier}), generate a group encryption key for the workspace, and invite other users (\S\ref{authorization}) to the workspace.

\subsection{Semantic User Identifiers}
\label{user-identifier}
As we have discussed in the above, having all users with their own semantic identifiers can facilitate user defined trust management. 
Unfortunately, users today, by and large, only possess in-app user identifiers assigned by
cloud-providers, such as email addresses, twitter IDs, or facebook IDs.
These in-app identifiers have two useful features: being globally unique and semantically meaningful (for example, friends can recognize each other by their email addresses or facebook IDs). 
Thus they can serve as Workspace user identifiers.
As a example, email servers provide unique user identifiers within their domain, and
an email address \name{alice@example.com} uniquely identifies the address owner Alice in the cyberspace.

Workspace assigns Alice an user identifier derived from Alice's email address.
As the first step of security bootstrapping a Workspace instance, Alice enters her email address \name{alice@example.com}, a personal key pair she uses for peer authentication and Invitation she received (\S\ref{authorization}).
Taking the Invitation, Workspace learns that \name{alice@example.com} is invited to workspace instance \name{MeetRoom}.
Therefore, Workspace first generates a new key pair, then names the name key with Alice assigned username \name{MeetRoom/alice@example.com}.
The \name{MeetRoom} key is used to sign Alice's data publications in this workspace instance.
Afterwards, following the naming convention defined in \S\ref{data-identifier}, Workspace uses Alice's personal key to issue a certificate for her \name{MeetRoom} key, so that whoever authenticates her personal key is able to verify her publications in \name{MeetRoom}.

Unlike traditional cloud-based applications, Workspace usernames are fully controlled by the users themselves.
Alice obtains the username and  ``MeetRoom'' key pair from her local Workspace application instance, without replying on external cloud-providers.

\subsection{Semantic Data Identifiers}
\label{data-identifier}
\begin{figure}[h]
    \centering
    \includegraphics[width=0.48\textwidth]{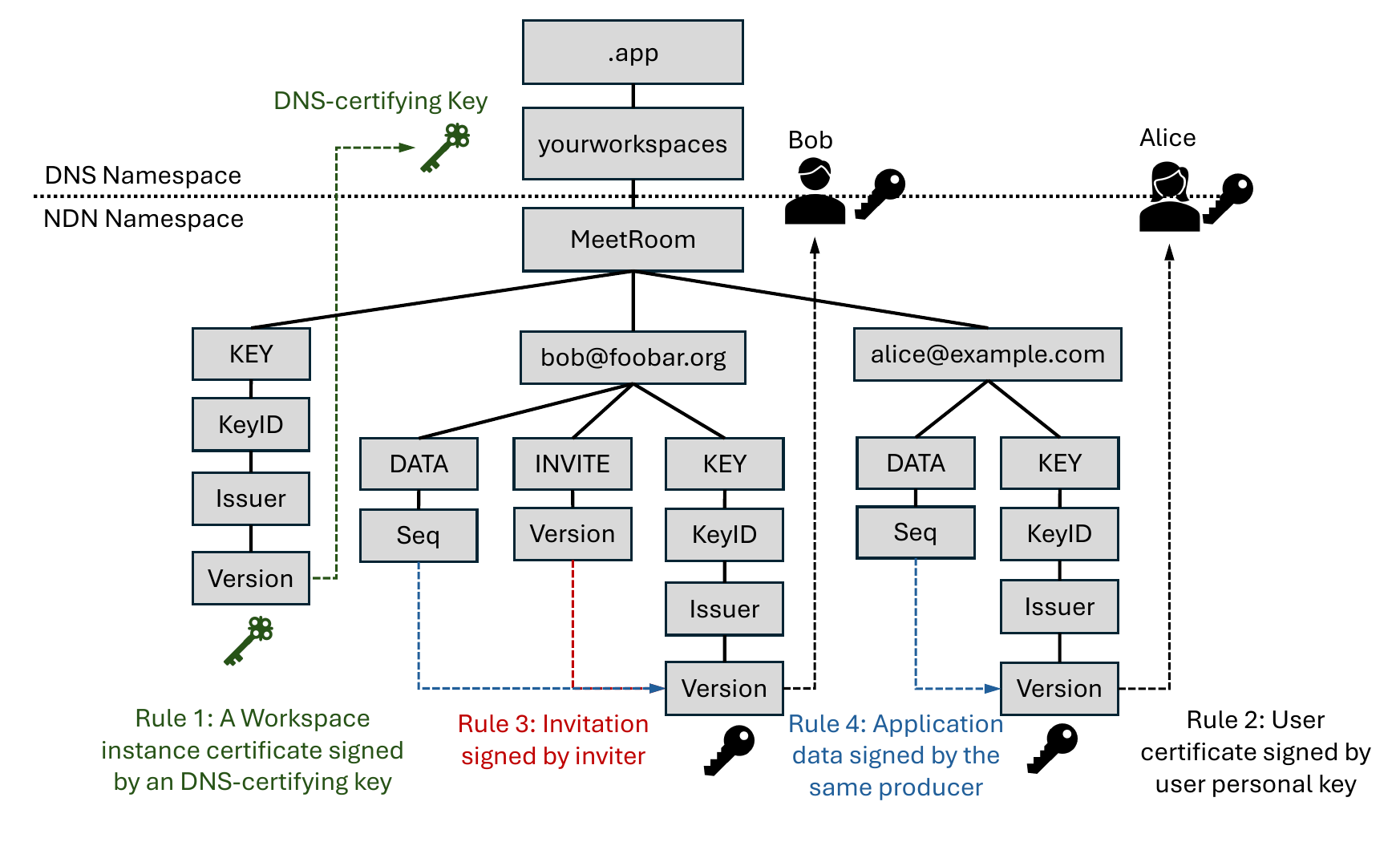}
    \caption{The Data namespace tree of ``MeetRoom'' workspace. Each node represents a name component. 
    }
    \label{fig:name-tree}
\end{figure}

Workspace semantically identifies each Data with a URL-like names.
Fig.~\ref{fig:name-tree} summarizes the Workspace Data namespace design when there are only Alice and Bob in \name{MeetRoom}.


Invitation Data (\S\ref{authorization}) are identified as \name{<workspace>/<inviter>/INVITE/<version>}, where \name{<workspace>} is workspace name, \name{inviter} indicates inviter user identifier who produces this Data, and \name{<version>} is the Invitation version.
For example, \name{MeetRoom/bob@foobar.org/INVITE/v=1} is an Invitation that \name{bob@foobar.org} sends to Alice.
Application Data follow the naming convention \name{/<workspace>/<username>/DATA/<seq>}, where \name{<username>} component is the user email address, and the \name{seq} indicates the sequence number of this user's publications in this workspace.
Therefore, Alice's first Data is named as \name{MeetRoom/alice@example.com/DATA/seq=1}.
User certificates are named with \name{<workspace>/<username>/KEY/<keyid>/<issuer>/<version>}.
The \name{<keyid>} is the key identifier of \name{<workspace>/<username>} key pair, and \name{<issuer>} indicates the certificate signer.
For instance, 
\name{MeetRoom/alice@example.com/KEY/0x1d4e5/self/v=1} would correspond to version one of the self-signed certificate owned by \name{alice@example.com}.

Workspace's schematized trust policies can be represented with signing relations between nodes on Data namespace tree.
Figure~\ref{fig:name-tree} shows four rules that \name{MeetRoom} workspace executes.
Rule 1 specifies the workspace instance certificate signed by the \name{yourworkspaces.app} domain key (\eg X.509 certificate), so that invitees can authenticate the instance name \name{MeetRoom}.
Rule 2 requires all user certificates signed by user's personal keys, and Rule 3 enforces all Invitation must be signed by the inviter itself, preventing \name{MeetRoom} user signing Invitations on behalf of others.
Rule 4 asks all data under \name{MeetRoom/<username>/DATA} prefix signed by the certificate of the same user.

\vspace{-2pt}
\subsection{Membership Management}
\label{authorization}
\begin{figure}
    \centering
    \includegraphics[width=0.48\textwidth]{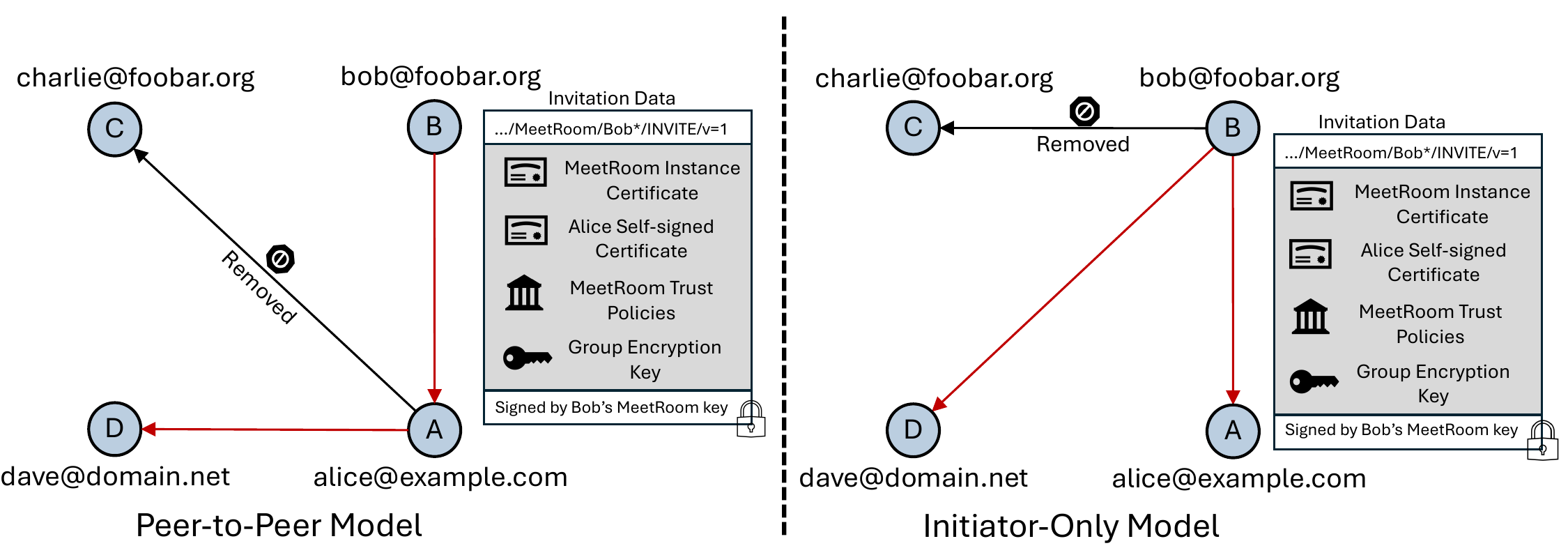}
    \caption{``MeetRoom'' workspace in two membership models.
    }
    \label{fig:member}
\end{figure}
Our initial NDN Workspace design supports two membership management models: \textit{initiator-only} and \textit{peer-to-peer}.
The former allows the workspace initiator (\ie the user who creates the workspace) to invite others to join the workspace, while the latter allows anyone in a workspace to invite others to join the collaboration.

As an example, in Fig.~\ref{fig:member}, we assume Bob and Alice are colleagues and know each other's self-signed certificate a prior. Bob initiates the \name{MeetRoom} workspace with the initiator-only model and invites Alice to join.
Bob cryptographically introduces her to the workspace by producing an \textit{Invitation} Data (discussed in~\S\ref{data-identifier}), which includes \name{MeetRoom} instance certificate, Alice's certificate for her personal key, the \name{MeetRoom} instance group encryption key (encrypted by Alice's public key) and trust policies, and finally sends the Invitation Data to Alice.

Accepting the Invitation, Alice executes the trust policies within it, which define the rules to validate Invitation and \name{MeetRoom} certificate.
Later when Bob introduces Charlie to the workspace, Alice and Charlie can securely communicate with each other since both personal keys are authenticated by Bob.
Each Invitation has its lifetime.
Bob can periodically renew Invitations and publish them to the workspace.
Alice is automatically removed if Bob decides not to renew Invitations.


In the peer-to-peer model, everyone can invite new members, Alice can invite Charlie following the same approach as Bob, and Charlie can invite other people too.
Each invitation introduces a valid certificate into the workspace.

\vspace{-2pt}
\subsection{Data Producing and Consuming Workflow}
\label{prod-cons}
When Alice makes a change in a document, her Workspace instance encrypts the changes and signs a Data packet.
To secure her data content, Workspace fetches the group encryption key from the storage module to encrypt named content (as shown in \S\ref{data-identifier}), uses her \name{MeetRoom} certificate to sign encrypted content, and produces a Data packet.

Receiving the published Data, Bob's instance validates Data by first
verifying the Data signature using Alice's certificate (as indicated in the KeyLocator field), then executing trust policies to check if the signing relation conforms to trust rules.
Validated Data is stored into Bob's browser storage, with document change applied to his local copy of the document.


\subsection{Synchronizing Documents}
\label{sync}

To ensure \name{MeetRoom} users have a consistent view on documents, Workspace instances need to synchronize on the latest data publication and merge concurrent updates from users. We use NDN Sync~(\S\ref{ndn}) for synchronizing the namespace to fetch these updates.

Specifically, we utilize the State Vector Sync (SVS)~\cite{tr-svs} protocol to synchronize the latest sequence numbers from each \name{MeetRoom} member. SVS uses multicast Sync Interests to notify all members in the workspace group about newly produced sequence numbers from other members.
Once a Workspace member learns a new sequence number, they can fetch the corresponding data by sending an Interest for the name, as described in \S\ref{data-identifier} and \S\ref{prod-cons}.

We note that SVS provides reliable namespace synchronization; Charlie may learn about Bob's data production from Alice, even when Bob is offline. Since data in Workspace is directly secured, security is independent from the channel through which it was fetched or where it was stored.
Therefore, participants (in this case, Alice) may also supply data from other producers (Bob) when the original producer is offline.


At the application layer, we utilize the Conflict-free Replicated Data Type~(CRDT) data structures for conflict resolution on concurrent changes from users.
Since SVS provides at-least-once delivery semantics, this guarantees eventual consistency between the users' view of the document.

\vspace{-3pt}
\subsection{Supporting Asynchronous Collaboration}
\label{async}

As data semantic (\S\ref{data-identifier}) and security (\S\ref{prod-cons}) are decoupled from communication channels, Workspace users can store their data anywhere.
Secured by Alice directly, Alice's data can be safely synchronized into every other user's local storage\footnote{Workspace uses Origin Private File System (OPFS) as persistent in-browser local storage.}, which enables Workspace supporting \textit{asynchronous} user-to-user communications. For example, if Alice makes changes when Bob is not available but Charlie is ``online'', Charlie will receive Alice's updates through documents synchronization (\S\ref{sync}).
Therefore when Charlie and Bob meet, Bob eventually receives Alice's update from Charlie.

In the fully asynchronous communication scenario where Alice makes update as the only ``online'' user, Workspace ensures data availability for Bob and Charlie using an NDN data repository (or \textit{Repo} for short) inside the network.
Repo is an in-network storage service~\cite{yupythonrepo}\footnote{Network providers should have multiple Repo instances to avoid single point of failure.} that can join workspace Sync groups, fetch latest Data publication inside the group, thus store all document changes in-network.
When the network receives an Interest asking for any Data that Repo already has, Repo directly replies the request with corresponding Data.

Additionally, Repo caches the latest incoming Sync Interests and it is aware of the latest state vectors in each workspace.
Whenever Repo receives a Sync Interest with outdated state vectors, Repo will send cached latest Sync Interests to the network, so that remote parties can learn the latest status of the Sync group.

Repo ensures that even if Alice makes changes when neither Bob or Charlie is available, Repo caches Alice's Sync Interest and fetches Alice's Data.
Later when Bob and Charlie become available and send Sync Interests to catch up with the latest status in the Sync group, they are informed with the state vector updated by Alice, and can fetch latest Data from Repo.

\section{Implementation}
\label{imple}
We have implemented NDN Workspace as a web application using 
TypeScript\footnote{\url{https://ndn-workspace.web.app/}}.
This section explains the implementation details of NDN Workspace, focusing on conflict resolution, data synchronization and local data storage.

\subsection{Conflict Resolution}
Each data produced by NDN Workspace carrying a user's update to the shared document. As a result, NDN Workspace needs to resolve conflicts from concurrent updates.
\begin{figure}[h]
\includegraphics[width=0.4\textwidth]{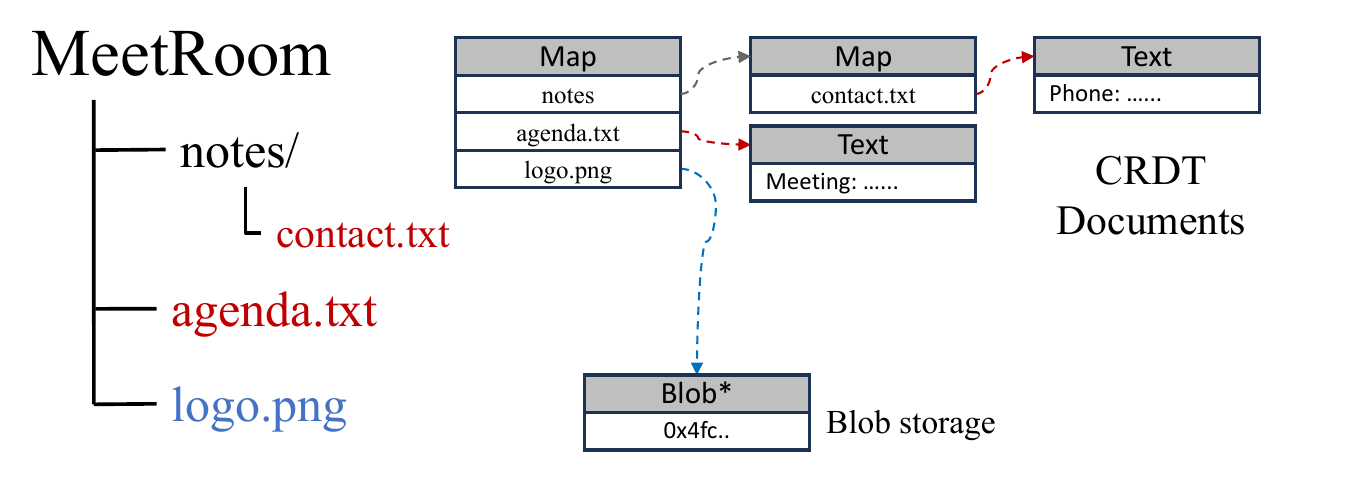} \caption{NDN Workspace represents file structure as a document, which is a collection of shared data structures.}\label{fig:crdt}
\end{figure}
\vspace{-5pt}
Our implementation automatically resolve conflicts by defining a document as a collection of Conflict-free Replicated Data Types (CRDT)~\cite{shapiro2011conflict}.
As shown in Fig.~\ref{fig:crdt}, we map workspace file structures to CRDT data types.
We represent folders as \textit{CRDT Maps}, which maintains a mapping between files and sub-structures.
Text files are mapped to \textit{CRDT Texts}, allowing real-time collaboration.
Other types of files, which do not support real-time collaboration, are transformed into immutable versioned, binary blobs, with CRDT only storing their data object names.
We utilized Yjs~\cite{yjs} as our CRDT implementation.

CRDT captures the updates in workspace and wraps them into Data objects and publish, as described in \S\ref{prod-cons}.
If the delta change is too large for an NDN Data packet, Workspace will wrap it into a separate segmented blob object, and put its Name in the Data object to publish.

\vspace{-5pt}
\subsection{Data Synchronization}
\label{rendezvous}
As described in \S\ref{sync}, our application utilizes NDN Sync to synchronize data publications among participants. Pragmatically, however, running Sync requires participants able to exchange NDN packets. 
We solve this problem by connecting up participants through the NDN Testbed network~\cite{ndn-testbed}.
NDN testbed network has about 25 router nodes across four continents, and is operated by collaborating universities and research institutes.
When starting up, each Workspace user connects to the geographically closest NDN router in the Testbed, and registers the user's prefixes to the router.
Workspace can then utilize Sync (\S\ref{sync}) to disseminate all updates among Workspace participants connected to the Testbed with no dependency on any third party rendezvous.


NDN Workspace users rendezvous through their shared application namespace; NDN Testbed network simply provides name-based data exchange among the users.
The Testbed routers neither understands the meaning of the names nor
parse the packets being forwarded; they simply performs bit-string matching between the names carried in the packets and their forwarding table (FIB).
Users or organizations may also set up their own NDN networks, and optionally connect these nodes to the existing NDN Testbed.
Traditional cloud servers, on the other hand, must understand the semantics of the underlying data to be able to route them to the correct clients, and thus need to be centrally controlled and secured.

It is also important to note that NDN Workspace, by running over NDN, can communicate over any communication media that support exchange packets by names. 
For example, NDN Workspace users can communicate directly over any layer-2 protocols.
\eg using direct WiFi on an airplane.
We also implemented the option of using PeerJS to establish direct $n\times n$ connectivity among all users in a group using WebRTC.

\section{Evaluation}
\label{evaluation}
In this section, we evaluate NDN Workspace from both performance and functionality perspective to show NDN Workspace's effectiveness and uniqueness compared to existing applications.

\subsection{Performance}
To demonstrate the effectiveness of our proof-of-concept implementation, we performed experiments to measure the delay for a user to receive a data publication.
More specifically, we ran multiple simulated users on the test machine, and let each simulated user randomly publish data carrying the timestamp of the publication.
When another simulated user received this publication, it logged the difference between reception and publication.

We conducted our experiment using an Ubuntu 22.04 machine with an AMD EPYC 7702P (\num{64} core, \qty{1.5}{\giga\hertz}) and \qty{256}{\giga\byte} memory.
We set the payload size of the publication to be \qty{100}{\byte}.
In the real-world scenario, the size of payload is dynamic, varying from several bytes (when a user is typing) to kilobytes (when a user pastes a large text block).
However, as stated later, data publication frequency is low enough in shared text document editing, our designated scenario.
Therefore, we can assume the payload will not affect performance significantly, as we always have enough bandwidth.

Data publication frequency also varies in different scenarios.
In a shared document scenario, the frequency is no more than user typing speed, which is average 160ms per keystroke~\cite{dhakal2018observations}.
There exists implementations\footnote{\url{https://github.com/overleaf/overleaf/blob/main/libraries/overleaf-editor-core/lib/ot_client.js}} that combine multiple keystrokes into one to improve performance.
NDN Workspace uses similar techniques to combine keystrokes, so that publication rate varies between per \qty{500}{\milli\second} and per \qty{1000}{\milli\second}.
For simplicity, we evaluated the scenario where simulated users are connecting to several NDN Testbed routers with data publication per second.

In this experiment, we had 16 simulated users, which is reasonable user number of a collaborative project.
All simulated users were in US and but in different geolocations.
we picked four NDN Testbed nodes as their closest NDN router, with one in west US, two in central US and one in east US.
The average latency was \qty{191.55}{\milli\second}.

\subsection{Functionality}

We are still in early stage of NDN Workspace trials, and yet to carry out systematic evaluation about NDN  Workspace's functionality.
As preliminary proof of evidence, in this subsection we first compare the functionalities with two typical text collaboration software in Table~\ref{tab:functionality}, then describe several anecdotes to show asynchronous collaboration, use of ad hoc connectivity, and the utility of NDN Workspace library in supporting other kinds of decentralized, collaborative applications.

\begin{table}[h]
	\scriptsize
	\resizebox{0.46\textwidth}{!}{
	\begin{tabular}{  
            > {\raggedright\arraybackslash} p{0.7cm}   
            > {\centering\arraybackslash} p{0.6cm}  
            > {\centering\arraybackslash} p{1.6cm} 
            > {\centering\arraybackslash} p{1.6cm} 
        }
	\toprule
		\multicolumn{1}{> {\centering\arraybackslash} p{0.7cm}}{\textbf{Software}} &
            \multicolumn{1}{> {\centering\arraybackslash} p{0.6cm}}{\textbf{Offline}} &
            \multicolumn{1}{> {\centering\arraybackslash} p{1.6cm}}{\textbf{Local Communication}} &
            \multicolumn{1}{> {\centering\arraybackslash} p{1.6cm}}{\textbf{Real-time Collaboration}} \\
    \midrule
		Overleaf\tablefootnote{We use Overleaf as a representative of cloud-based collaborative editors.} & \ding{54} & \ding{54} & \ding{52} \\
            Git & \ding{52} & \ding{54} & \ding{54}  \\
            Workspace & \ding{52} & \ding{52} & \ding{52}  \\
    \bottomrule
	\end{tabular}
	}
	\smallskip
	\caption{Functionality Comparison}
	\label{tab:functionality}
\end{table}

\noindent\textbf{Offline support}: Though today's Internet coverage is high, there are still some cases when Internet is not available. For example, when travelling in a plane, Internet access may be expensive.
Both Git and  NDN Workspace store user data locally, so that user can edit shared documents when the Internet is not available or intermittent.
Overleaf, on the other hand, requires establishing connections to servers.

\noindent\textbf{Local Communication}:  NDN Workspace does not only offer offline availability, but also allows users to collaborate locally by exploiting local connectivity.
Since NDN is a network layer protocol which can run over any types of packet transport ranging from TCP/UDP tunnels to Ethernet, WiFi or Bluetooth, NDN Workspace is able to make use of all available connectivity.
For example, users on an airplane can establish NDN connectivity over Bluetooth, or directed WiFi.
An enterprise can also provide local NDN node as a rendezvous point or a WebRTC server to setup peer-to-peer connections.

\noindent\textbf{Real-time Collaboration}: Traditional local software, such as Git, often leads to conflict when multiple users are editing the same file simultaneously.
The conflicts has to be resolved manually in Git.
Overleaf (and other cloud-based collaborative editors) allows automatic conflict resolution using Operational Transformation (OT), but requires users to connect to the server.
NDN Workspace, on the other hand, makes use of CRDT to resolve conflicts, enabling asynchronous real-time collaboration.
Two users reach consensus state as long as they receive the same set of delta updates, no matter what is the order of reception.


\noindent\textbf{A1: Collaboration from Anywhere.}
We have used NDN Workspace in several NDN community events.
We used it for NDN Retreat in late 2023 and NDN Community Meeting in March 2024, inviting people to jointly edit files and share presentations with other onsite or remote attendees connected through the NDN Testbed network.
Since February 2024, we have been using NDN Workspace in our research group to log  research progress and report issues through shared files.
People edit shared files independently from whether they are online or offline, and if offline, the changes get synchronized once one gets connected.




\noindent\textbf{A2: Exchanging Named Data over Any Connectivity.}
Also in February 2024, two members of the group attended a technical conference.
During the long flight back home, the two used NDN Workspace to jointly edit the NDN Workspace design document (hosted in our research group workspace instance) via one member's WiFi hotspot, which only has no Internet access.
It shows that by securing data directly, NDN Workspace is able to make use any connectivity that can exchange named data.

\noindent\textbf{A3: Developing Applications with NDN Workspace Libraries.}
During the development of NDN Workspace, we built up a set of 
libraries \footnote{\url{https://github.com/UCLA-IRL/ndnts-aux} provides the building blocks for NDN Workspace, which focuses on providing web front-end and system assembly.} for others to make use of NDN Workspace design and develop their own applications.
A developer who was new to the NDN Workspace design implemented a simple decentralized chat app \footnote{The code can be accessed at https://ndn-workspace.web.app/} over a short few weeks by making use of the NDN Workspace library.
\vspace{-5pt}
\section{Discussion}
\label{discuss}

In this section, we step up a level to articulate the major lessons we learned from the NDN Workspace development, the viability of using semantic identifiers in networking and applications, and what outcome from the NDN Workspace development could be utilized in developing future metaverse applications.

\noindent\textbf{What we learned from NDN Workspace development}:
A mandatory requirement in networking today is that all communications be secured. 
The first lesson we learned from developing NDN Workspace, and from the NDN research over the last decade, is that security starts with semantic identities. 
Cryptographic signatures and data encryptions, by themselves, are not security but tools to execute security policies whose definition requires semantic identifiers.
To have security designed into NDN Workspace, its design uses semantic naming for both users and data, enabling NDN Workspace to support user authentications and to secure data directly.

It is also well known that web based applications are data-oriented, where individual entities request data from others and supply its own data production to others. 
The second lesson we learned is that data-centric networking, in replacement of node/connection-centered networking, is a great fit to distributed applications. 
It enables individual entities to exchange named data directly.
Application Data Units (ADUs) is a concept introduced by Clark~\cite{clark90} over three decades ago.  NDN extends this concept by assigning each ADU a semantic name, and securing ADUs directly, enabling exchanging data objects and enforcing fine-grained access control in a distributed application.

Communication by named, secured data object exchanges opens the door to supporting asynchronous applications where participants may not be all online all the time. 
Securing data directly removes the dependency on trusting specific nodes, so that data may come from any nodes. As the third lesson we learned, data-centric networking allows app-independent storage in the network, which in turn enables NDN Workspace to provide support for asynchronous collaborative editing.

\noindent\textbf{The Viability of Using Semantic Identifiers}:
All the functionalities mentioned above lie on assigning users and data unique and semantic identifiers, which are the foundation to security.
A common question about taking on this approach is where to obtain such identifiers.
We would like to point out that, networked applications today, by and large, are already built by using semantic identifiers derived from DNS names, ranging from old email to more recent instagram and WhatsApp (each of them identifies an application). 
Taking advantage of DNS's hierarchical namespace structure with delegations, 
the NDN Workspace design follows the same approach taken by these well known apps, and further extends the DNS namespace to identify users and data with fine granularity.
We hope that NDN Workspace sets an example in promoting the use of DNS-derived semantic naming in future application development, enabling them to have security \emph{designed in}, instead of patching security onto the communication channels.

\noindent\textbf{New Insights to Internet Decentralization}:
As pointed by \cite{huitema2023report}, the one important contributing factor of Internet centralization and consolidation is lack of security in networking, which pushed defenses to higher layers (\eg HTTPS) and mandated cloud-based solutions.
This work brings new insight to develop decentralized applications by proposing a new networking platform of applications with security built-in.

\noindent\textbf{Developing New Metaverse Apps over NDN}\quad 
NDN Workspace shows an example of developing distributed and decentralized applications by having users directly exchange semantically named and secured data.
We believe that the same approach can be applied to other metaverse applications.
We summarise below a few worth-noting lessons we learned from building NDN Workspace.
\begin{enumerate}
    \item Designing a good \textit{semantic naming scheme} is a cornerstone in developing a data-centric application. Once we name all pieces of data semantically, the application can utilize existing NDN primitives such as Sync for communication and security libraries for security policies and access control.

    \item Our design utilizes email addresses to identify users; other existing internet identifiers, such as DNS names, can be used in similar way. The only requirement is that these identifiers should be unique and semantically meaningful.
    We are well aware that people may have privacy concerns on using semantic user identifiers~\cite{zhang2021investigating}, an issue we will address in the future work.
    
    \item NDN enables the concept of a generic in-network storage, data repository (repo), that can be considered as part of \textit{network infrastructure} support together with network connectivity, to enable asynchronous data exchanges among end users. NDN secures data directly, thus decoupling application security from the trust on specific nodes. As a result, a generic data repository can \textit{securely} store data for applications, without understanding, parsing or decrypting the data itself. Applications can subsequently fetch data by names and verify the authenticity of the data.
\end{enumerate}

\begin{figure}[h]
\includegraphics[width=0.4\textwidth]{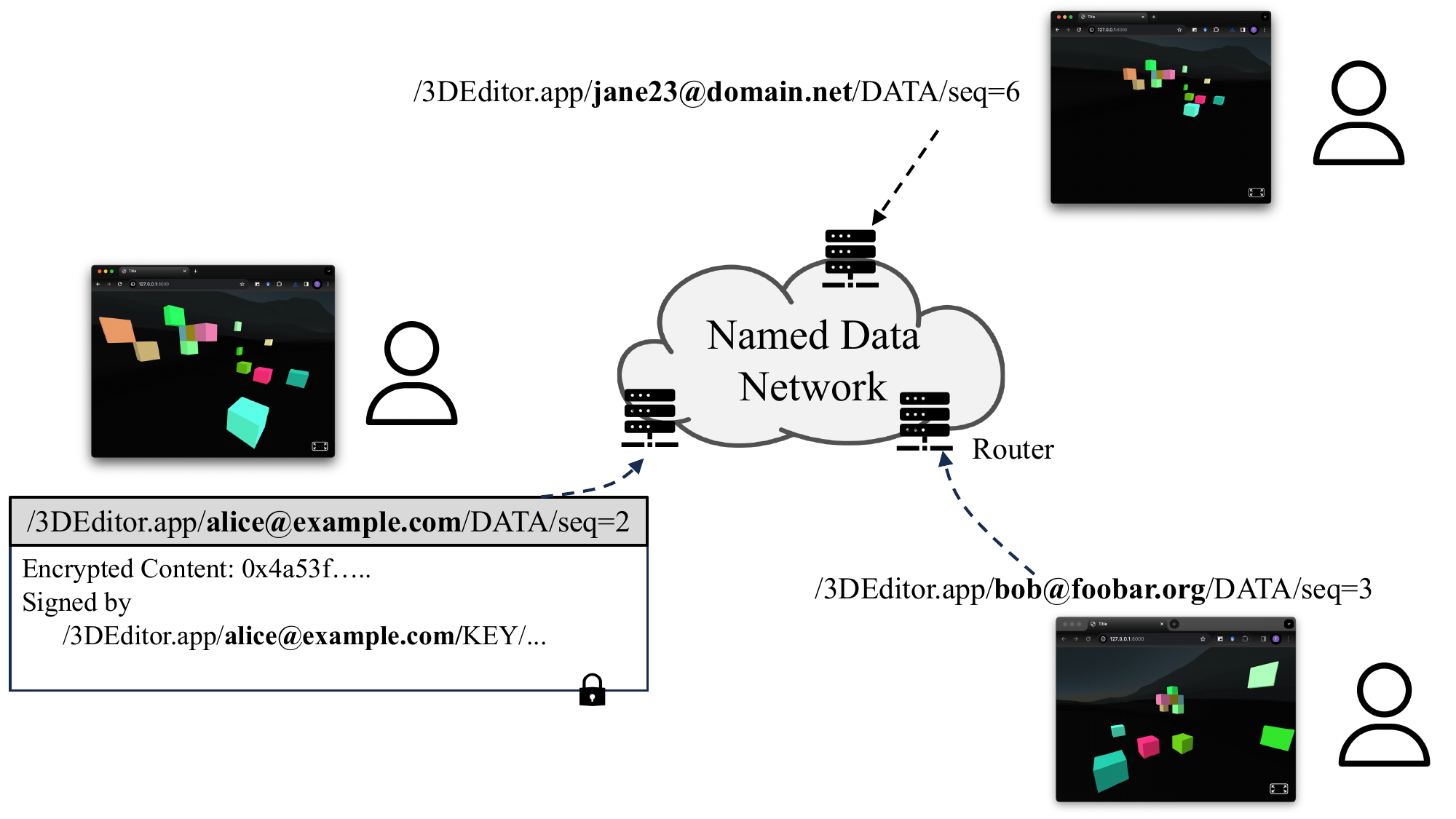} \caption{A collaborative 3D editing scenario among Alice,
Bob and Jane under namespace \texttt{/3DEditor.app}}\label{fig:microverse-design}
\vspace{5pt}
\end{figure}

Based on the lessons learned, we also developed a prototype of shared 3D~scene editor, enabling collaborative editing of objects in 3D scenes expressed in JSON documents~\cite{microverse}. As a demonstrative metaverse application, the 3D~editor allows placing objects into the scene, and uses Sync to synchronize the objects in realtime as multiple users manipulate them. 
As Fig.~\ref{fig:microverse-design} shows, scene updates are carried in Data objects with URI-like names, and each user secures both asset and update objects by signing the object using their certificate.

We note that since this is not a security paper, we did not specifically discuss the threat model, as we focus more on our application design. Such a detailed analysis can be considered as part of our future work.

\section{Conclusion and Remaning Works}
\label{conclude}

As we show in this paper, the NDN Workspace design is built on a solid foundation provided by Named Data Networking (NDN) with its decade-long research results.
We believe that the same approach can be applied to developing new metaverse applications, enabling them to exchange semantically named data objects with security built-in, and supporting fine granularity access control by defined trust policies using names. The development of NDN Workspace can serve as a starting point for developing more advanced metaverse applications over NDN. 
We hope that the primitives and libraries developed in support of the NDN Workspace implementation could be readily utilized for building such applications. 

We also identify a few remaining tasks.
First, we need to further investigate NDN Workspace membership management (\S\ref{authorization}) and practice the design of peer-to-peer mode.
Second, we will improve the implementation of document synchronization (\S\ref{sync}).
CRDT algorithms share similarities with SVS (\S\ref{ndn}) on state vectors and logical clocks.
Current NDN Workspace implementation treats CRDT as a blackbox and uses the high-level CRDT APIs to resolve conflict conflicts.
We will explore CRDT algorithms details and make use of existing libraries more efficiently.
Third, we also plan to further mature NDN Workspace design based on feedback from users, in particular in providing user-friendly interface for defining trust policies, plowing through the way towards this new direction of designing collaborative applications by exchanging semantically named and secured data. 



\bibliographystyle{IEEEtran}
\bibliography{reference}

\begin{thebibliography}{10}
\providecommand{\url}[1]{#1}
\csname url@samestyle\endcsname
\providecommand{\newblock}{\relax}
\providecommand{\bibinfo}[2]{#2}
\providecommand{\BIBentrySTDinterwordspacing}{\spaceskip=0pt\relax}
\providecommand{\BIBentryALTinterwordstretchfactor}{4}
\providecommand{\BIBentryALTinterwordspacing}{\spaceskip=\fontdimen2\font plus
\BIBentryALTinterwordstretchfactor\fontdimen3\font minus \fontdimen4\font\relax}
\providecommand{\BIBforeignlanguage}[2]{{%
\expandafter\ifx\csname l@#1\endcsname\relax
\typeout{** WARNING: IEEEtran.bst: No hyphenation pattern has been}%
\typeout{** loaded for the language `#1'. Using the pattern for}%
\typeout{** the default language instead.}%
\else
\language=\csname l@#1\endcsname
\fi
#2}}
\providecommand{\BIBdecl}{\relax}
\BIBdecl

\bibitem{ndn14}
L.~Zhang, A.~Afanasyev, J.~Burke, V.~Jacobson, k.~claffy, P.~Crowley, C.~Papadopoulos, L.~Wang, and B.~Zhang, ``{Named Data Networking},'' \emph{ACM SIGCOMM Computer Communication Review (CCR)}, vol.~44, no.~3, pp. 66--73, Jul. 2014.

\bibitem{2021plug-play}
\BIBentryALTinterwordspacing
T.~Yu, P.~Moll, Z.~Zhang, A.~Afanasyev, and L.~Zhang, ``Enabling plug-n-play in named data networking,'' in \emph{MILCOM 2021 - 2021 IEEE Military Communications Conference (MILCOM)}.\hskip 1em plus 0.5em minus 0.4em\relax IEEE Press, 2021. [Online]. Available: \url{https://doi.org/10.1109/MILCOM52596.2021.9653033}
\BIBentrySTDinterwordspacing

\bibitem{yu2021enabling}
------, ``Enabling plug-n-play in named data networking,'' in \emph{MILCOM 2021-2021 IEEE Military Communications Conference (MILCOM)}.\hskip 1em plus 0.5em minus 0.4em\relax IEEE, 2021, pp. 562--569.

\bibitem{yu2015schematizing}
\BIBentryALTinterwordspacing
Y.~Yu, A.~Afanasyev, D.~Clark, k.~claffy, V.~Jacobson, and L.~Zhang, ``Schematizing trust in named data networking,'' in \emph{Proceedings of the 2nd ACM Conference on Information-Centric Networking}, ser. ACM-ICN '15.\hskip 1em plus 0.5em minus 0.4em\relax New York, NY, USA: Association for Computing Machinery, 2015, p. 177–186. [Online]. Available: \url{https://doi.org/10.1145/2810156.2810170}
\BIBentrySTDinterwordspacing

\bibitem{nichols2021trustschemas}
\BIBentryALTinterwordspacing
K.~Nichols, ``Trust schemas and icn: key to secure home iot,'' in \emph{Proceedings of the 8th ACM Conference on Information-Centric Networking}, ser. ICN '21.\hskip 1em plus 0.5em minus 0.4em\relax New York, NY, USA: Association for Computing Machinery, 2021, p. 95–106. [Online]. Available: \url{https://doi.org/10.1145/3460417.3482972}
\BIBentrySTDinterwordspacing

\bibitem{yu2023trustschema}
T.~Yu, X.~Ma, H.~Xie, Y.~Kocaoğullar, and L.~Zhang, ``A new api in support of ndn trust schema,'' 10 2023, pp. 46--54.

\bibitem{2018nac}
Z.~Zhang, Y.~Yu, S.~K. Ramani, A.~Afanasyev, and L.~Zhang, ``Nac: Automating access control via named data,'' in \emph{MILCOM 2018 - 2018 IEEE Military Communications Conference (MILCOM)}, 2018, pp. 626--633.

\bibitem{syncsok}
\BIBentryALTinterwordspacing
P.~Moll, V.~Patil, L.~Wang, and L.~Zhang, ``Sok: The evolution of distributed dataset synchronization solutions in ndn,'' in \emph{Proceedings of the 9th ACM Conference on Information-Centric Networking}, ser. ICN '22.\hskip 1em plus 0.5em minus 0.4em\relax New York, NY, USA: Association for Computing Machinery, 2022, p. 33–44. [Online]. Available: \url{https://doi.org/10.1145/3517212.3558092}
\BIBentrySTDinterwordspacing

\bibitem{tr-svs}
P.~Moll, V.~Patil, N.~Sabharwal, and L.~Zhang, ``{A Brief Introduction to State Vector Sync},'' NDN, Technical Report NDN-0073, Apr. 2021.

\bibitem{yupythonrepo}
T.~Yu, Z.~Kong, X.~Ma, L.~Wang, and L.~Zhang, ``Pythonrepo: Persistent in-network storage for named data networking.''

\bibitem{shapiro2011conflict}
M.~Shapiro, N.~Pregui{\c{c}}a, C.~Baquero, and M.~Zawirski, ``Conflict-free replicated data types,'' in \emph{Stabilization, Safety, and Security of Distributed Systems: 13th International Symposium, SSS 2011, Grenoble, France, October 10-12, 2011. Proceedings 13}.\hskip 1em plus 0.5em minus 0.4em\relax Springer, 2011, pp. 386--400.

\bibitem{yjs}
\BIBentryALTinterwordspacing
K.~Jahns, ``Yjs: Shared data types for building collaborative software,'' 2024. [Online]. Available: \url{https://yjs.dev/}
\BIBentrySTDinterwordspacing

\bibitem{ndn-testbed}
T.~N. Team, ``Ndn testbed,'' Online at https://named-data.net/ndn-testbed/, 2024.

\bibitem{dhakal2018observations}
V.~Dhakal, A.~M. Feit, P.~O. Kristensson, and A.~Oulasvirta, ``Observations on typing from 136 million keystrokes,'' in \emph{Proceedings of the 2018 CHI conference on human factors in computing systems}, 2018, pp. 1--12.

\bibitem{clark90}
\BIBentryALTinterwordspacing
D.~D. Clark and D.~L. Tennenhouse, ``Architectural considerations for a new generation of protocols,'' in \emph{Proceedings of the ACM Symposium on Communications Architectures \&amp; Protocols}, ser. SIGCOMM '90.\hskip 1em plus 0.5em minus 0.4em\relax New York, NY, USA: Association for Computing Machinery, 1990, p. 200–208. [Online]. Available: \url{https://doi.org/10.1145/99508.99553}
\BIBentrySTDinterwordspacing

\bibitem{huitema2023report}
C.~Huitema, G.~Huston, D.~Kutscher, and L.~Zhang, ``Report of 2021 dinrg workshop on centralization in the internet,'' \emph{ACM SIGCOMM Computer Communication Review}, vol.~53, no.~2, pp. 31--39, 2023.

\bibitem{zhang2021investigating}
Z.~Zhang, S.~Y. Won, and L.~Zhang, ``Investigating the design space for name confidentiality in named data networking,'' in \emph{MILCOM 2021-2021 IEEE Military Communications Conference (MILCOM)}.\hskip 1em plus 0.5em minus 0.4em\relax IEEE, 2021, pp. 570--576.

\bibitem{microverse}
\BIBentryALTinterwordspacing
J.~Burke, L.~Zhang, and D.~Kutscher, ``Named data microverse,'' Mar 2023. [Online]. Available: \url{https://named-data.net/microverse/}
\BIBentrySTDinterwordspacing

\end{thebibliography}


\end{document}